\documentclass[a4paper,fleqn,usenatbib]{mnras}

\usepackage{newtxtext,newtxmath}

\usepackage[T1]{fontenc}
\usepackage{ae,aecompl}

\usepackage{graphicx}	
\usepackage{amsmath}	
\usepackage{amssymb}	

\usepackage{upgreek}
\usepackage{hyperref}
\usepackage{breakurl}
\usepackage{relsize}
\renewcommand{\ion}[2]{\ensuremath{\text{#1\hspace{0.15em}{\smaller #2}}}}
\renewcommand{\vec}[1]{\ensuremath{\boldsymbol{#1}}}

\newcommand{\lognhi}[0]{\ensuremath{\log_{10}(N_{\ion{H}{I}} / \mathrm{cm}^{-2})}}
\newcommand{\vel}[1]{\ensuremath{v_{\rm #1} / (\mathrm{km \, s}^{-1})}}
\newcommand{\scalprod}[0]{\ensuremath{\boldsymbol{\cdot}}}
\newcommand{\software}[1]{\textsc{\MakeLowercase{#1}}}

\title[All-sky map of high-velocity clouds]{A new all-sky map of Galactic high-velocity clouds from the 21-cm HI4PI survey}

\author[T.~Westmeier]{
Tobias Westmeier$^{1}$\thanks{E-mail: tobias.westmeier@uwa.edu.au} \\
$^{1}$International Centre for Radio Astronomy Research (ICRAR), The University of Western Australia, 35 Stirling Highway,\\ Crawley WA 6009, Australia
}

\date{Accepted XXX. Received YYY; in original form ZZZ}

\pubyear{2017}

\begin{document}
\label{firstpage}
\pagerange{\pageref{firstpage}--\pageref{lastpage}}
\maketitle

\begin{abstract}
	High-velocity clouds (HVCs) are neutral or ionised gas clouds in the vicinity of the Milky Way that are characterised by high radial velocities inconsistent with participation in the regular rotation of the Galactic disc. Previous attempts to create a homogeneous all-sky \ion{H}{I} map of HVCs have been hampered by a combination of poor angular resolution, limited surface brightness sensitivity and suboptimal sampling. Here, a new and improved \ion{H}{I} map of Galactic HVCs based on the all-sky HI4PI survey is presented. The new map is fully sampled and provides significantly better angular resolution ($16.2$ versus $36~\mathrm{arcmin}$) and column density sensitivity ($2.3$ versus $3.7 \times 10^{18}~\mathrm{cm}^{-2}$ at the native resolution) than the previously available LAB survey. The new HVC map resolves many of the major HVC complexes in the sky into an intricate network of narrow \ion{H}{I} filaments and clumps that were not previously resolved by the LAB survey. The resulting sky coverage fraction of high-velocity \ion{H}{I} emission above a column density level of $2 \times 10^{18}~\mathrm{cm}^{-2}$ is approximately 15~per cent, which reduces to about 13~per cent when the Magellanic Clouds and other non-HVC emission are removed. The differential sky coverage fraction as a function of column density obeys a truncated power law with an exponent of $-0.93$ and a turnover point at about $5 \times 10^{19}~\mathrm{cm}^{-2}$. \ion{H}{I} column density and velocity maps of the HVC sky are made publicly available as FITS images for scientific use by the community.
\end{abstract}

\begin{keywords}
	ISM: clouds -- Galaxy: halo -- Galaxy: kinematics and dynamics -- radio lines: ISM
\end{keywords}



\section{Introduction }
\label{sec:intro}

High-velocity clouds (HVCs) are gas clouds detected in the optical and ultraviolet, but most notably in the 21-cm line of neutral hydrogen, across much of the sky at radial velocities that are incompatible with the regular rotation of the Galactic disc. There is no general consensus on how to separate HVCs from gas at low and intermediate velocities. In the past, most authors applied a fixed velocity threshold of $90$ or $100~\mathrm{km \, s}^{-1}$ in the Local Standard of Rest (LSR). However, this is clearly insufficient, as Galactic disc emission still occupies such extreme velocities in some parts of the sky. An improved definition was proposed by \citet{Wakker1991a} who introduced a so-called \emph{deviation velocity} of $v_{\rm dev} = 50~\mathrm{km \, s}^{-1}$ to characterise HVCs as deviating by a fixed velocity separation from the maximally permissible velocity of Galactic disc emission in a given direction. Today, the concept of deviation velocity is the most commonly applied criterion for defining high-velocity gas, although the actual value of $v_{\rm dev}$ may differ from the one applied by \citet{Wakker1991a}.

HVCs come in a wide range of sizes and shapes ranging from large complexes and streams, such as the Magellanic Stream \citep{Mathewson1974} or complex~C \citep{Hulsbosch1968}, to (ultra-)compact and isolated HVCs (CHVCs/UCHVCs; \citealp{Braun1999,Adams2013}). The discovery of 21-cm \ion{H}{I} emission from HVCs by \citet{Muller1963} sparked an intense scientific debate about their spatial distribution and origin \citep[e.g.,][]{Oort1966}. This debate was settled only recently when distance brackets or upper limits for several HVCs became available, most notably halo gas in the direction of the Large Magellanic Cloud \citep{Savage1981}, complex~M \citep{Danly1993}, complex~A \citep{Wakker1996,vanWoerden1999}, complex~C \citep{Wakker2007,Thom2008}, the Cohen Stream \citep{Wakker2008}, complex~GCP \citep{Wakker2008} and complex~WD \citep{Peek2016}, confirming that HVCs are generally located several kpc above the Galactic plane in close proximity to the Milky Way. This rules out the hypothesis that HVCs are primordial dark-matter haloes distributed throughout the Local Group \citep{Blitz1999}, instead suggesting diverse gas infall or outflow mechanisms as the origin of most of the clouds \citep{Cox1972,Bregman1980,deBoer2004,Lehner2010,Fraternali2015,Fox2016,Marasco2017}. HVCs also contain significant components of ionised gas that can be traced with the help of optical and ultra-violet absorption spectroscopy \citep{Sembach2003,Fox2006}. For a comprehensive review of the properties and origin of HVCs the reader is referred to \citet{Wakker1997a}, \citet{Wakker2001} and \citet{vanWoerden2004}.

The issue of creating complete all-sky maps of HVCs has been strongly tied to the availability of homogeneous, sensitive, all-sky surveys of Galactic \ion{H}{I} emission. An initial, rather patchy all-sky map of HVCs based on the combination of different data sources was presented by \citet{Bajaja1985}. The map included data from the IAR 30-m antenna and already showed the general outline of several of the large HVC complexes known today, including the Magellanic Stream and Leading Arm, complex~C and the anti-centre complex. A much improved HVC survey of the northern sky with the 25-m Dwingeloo radio telescope was presented by \citet{Hulsbosch1988}. Their map reveals all of the large HVC complexes in the northern sky, including complexes~A, C and~M as well as the entire anti-centre complex. \citet{Wakker1991a} then combined the southern-hemisphere data from \citet{Bajaja1985} with the improved northern-hemisphere data from \citet{Hulsbosch1988} to create an all-sky HVC map which revealed for the first time the structure and kinematics of high-velocity gas across the entire sky, albeit at a fairly low angular resolution (also see \citealp{Wakker1997a}).

The possibilities of imaging HVCs across the entire sky were significantly improved by the new generation of sensitive \ion{H}{I} surveys carried out towards the end of the 20$^{\rm th}$ century, particularly the Leiden/Dwingeloo Survey (LDS; \citealp{Hartmann1997}), the IAR 30-m survey of the southern sky \citep{Arnal2000,Morras2000} and the \ion{H}{I} Parkes All-Sky Survey (HIPASS; \citealp{Barnes2001}). Despite being an extragalactic survey, HIPASS was successfully employed by \citet{Putman2002} to create an improved catalogue of HVCs in the southern hemisphere at much better angular resolution of $15.5~\mathrm{arcmin}$, but poor velocity resolution of $26.4~\mathrm{km \, s}^{-1}$. Similarly, \citet{deHeij2002a} extracted a catalogue of CHVCs from the LDS data. Both catalogues were merged by \citet{deHeij2002b} into an all-sky catalogue of CHVCs.

\begin{figure}
	\centering
	\includegraphics[clip,width=\linewidth]{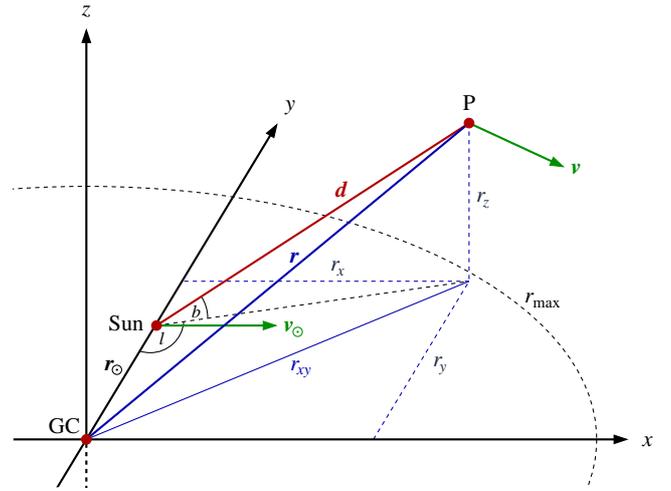}
	\caption{Illustration of the three-dimensional coordinate system and geometry assumed in the model of the Galactic disc, where an arbitrary point P with Galactic coordinates $(l, b)$ is seen from the position of the Sun along the line-of-sight vector $\vec{d}$. GC denotes the Galactic centre.}
	\label{fig:geometry}
\end{figure}

The LDS and IAR surveys were combined into the all-sky Leiden/Argentine/Bonn survey (LAB; \citealp{Kalberla2005}) which, in turn, was used by \citet{Westmeier2007} to create a homogeneous all-sky map of HVCs at an angular resolution of about $36~\mathrm{arcmin}$. While this map covers the full sky at the same resolution and sensitivity for the first time, the poor angular resolution of the data does not sufficiently resolve the most compact HVCs as well as the small-scale structure within larger HVC complexes.

The purpose of this paper is to present an improved, homogeneous all-sky map of HVCs based on the new HI4PI all-sky ($4 \uppi~\mathrm{sr}$) \ion{H}{I} survey \citep{BenBekhti2016}. The HI4PI survey combines data from the Effelsberg--Bonn \ion{H}{I} Survey (EBHIS; \citealp{Kerp2011}) in the northern hemisphere and the Galactic All-Sky Survey (GASS; \citealp{McClureGriffiths2009}) in the southern hemisphere to create an all-sky survey of Galactic \ion{H}{I} emission at a significantly better angular resolution of $16.2~\mathrm{arcmin}$. This results in a factor of $2.2$ improvement in resolution -- or a factor of about $5$ in beam solid angle -- compared to the previously available HVC map by \citet{Westmeier2007} based on the LAB survey. The resulting HVC map reveals details on much smaller angular and physical scales than before. Selected data products from the new map have been made publicly available, and their use and scientific exploitation by the community are encouraged.

This paper is organised as follows. In Section~\ref{sect_hi4pi} the basic properties of the HI4PI survey are briefly described. Section~\ref{sect_processing} explains the masking of Galactic \ion{H}{I} emission and the generation of HVC maps based on a simple model of the Milky Way. Section~\ref{sect_discussion} discusses the resulting all-sky HVC map, including a comparison with the previous map from \citet{Westmeier2007} based on the LAB survey. Lastly, the publicly released data products are introduced in Section~\ref{sect_release} followed by a brief summary in Section~\ref{sect_summary}.

\section{The HI4PI survey}
\label{sect_hi4pi}

The HI4PI survey \citep{BenBekhti2016} combines \ion{H}{I} data of the northern hemisphere from EBHIS \citep{Kerp2011,Winkel2016} obtained with the 100-m Effelsberg radio telescope with comparable data of the southern hemisphere from GASS \citep{McClureGriffiths2009,Kalberla2010,Kalberla2015} taken with the 64-m Parkes radio telescope. The GASS data were spectrally smoothed to the velocity resolution of EBHIS ($\Delta v \approx 1.5~\mathrm{km \, s}^{-1}$) while the EBHIS data were smoothed to the angular resolution of GASS ($\vartheta \approx 16.2~\mathrm{arcmin}$) to form a homogeneous all-sky survey of Galactic \ion{H}{I} emission with a sensitivity of $\sigma_{\rm RMS} \approx 43~\mathrm{mK}$. The latter corresponds to a $5 \, \sigma_{\rm RMS}$ column density sensitivity of $N_{\ion{H}{I}} \approx 2.3 \times 10^{18}~\mathrm{cm}^{-2}$, integrated over $20~\mathrm{km \, s}^{-1}$, for emission filling the beam. The LSR velocity coverage is about $\pm 480~\mathrm{km \, s}^{-1}$ in the southern hemisphere and $\pm 600~\mathrm{km \, s}^{-1}$ in the northern hemisphere. The HVC map presented here is restricted to the narrower velocity coverage of the GASS data to ensure a homogeneous coverage across the sky. As virtually all HVC emission is restricted to velocities of $|v_{\rm LSR}| < 450~\mathrm{km \, s}^{-1}$, the chosen velocity cut-off is not expected to have any effect on the quality and completeness of the HVC map.

\section{Data processing}
\label{sect_processing}

\subsection{Model of the Galactic \ion{H}{I} disc}

Creation of an all-sky HVC map requires masking of \ion{H}{I} emission from the Galactic disc. Hence, the first step will be to develop a three-dimensional model of the Milky Way's neutral gas disc. For the sake of simplicity, a simple cylindrical geometry of the Galactic disc is assumed, similar to the approach by \citet{Wakker1991a} who concluded that the purity of the resulting HVC map is not particularly sensitive to the details of the model. The basic Galactocentric coordinate system used here is illustrated in Fig.~\ref{fig:geometry}, where the $x$ and $y$ axes define the Galactic plane, while the $z$ axis is perpendicular to the plane and pointing in the direction of the north Galactic pole. The Sun is assumed to be located in the Galactic plane at the position of $\vec{r}_{\sun} = (0, r_{\sun}, 0)$ and moving in positive $x$-direction at a velocity of $\vec{v}_{\sun} = (v_{\sun}, 0, 0)$.\footnote{Strictly speaking, this velocity refers to the LSR, as the Sun possesses a small peculiar motion with respect to the LSR.} Throughout this paper the standard values of $r_{\sun} = 8.5~\mathrm{kpc}$ and $v_{\sun} = 220~\mathrm{km \, s}^{-1}$ adopted by the International Astronomical Union \citep{Kerr1986} are assumed.

Let us further introduce a point $\mathrm{P}$ with radius vector $\vec{r} = (r_{x}, r_{y}, r_{z})$ at which the gas is moving at a velocity of $\vec{v} = (v_{x}, v_{y}, 0)$. Following the geometry in Fig.~\ref{fig:geometry}, the position of $\mathrm{P}$ is given as
\begin{equation}
	\vec{r} = \begin{pmatrix} r_{x} \\ r_{y} \\ r_{z} \end{pmatrix} = \begin{pmatrix} d \sin(l) \cos(b) \\ r_{\sun} - d \cos(l) \cos(b) \\ d \sin(b) \end{pmatrix} \label{eqn:radius}
\end{equation}
where $l$ and $b$ are the Galactic longitude and latitude, respectively, of the line-of-sight vector, $\vec{d}$, pointing from the Sun to point $\mathrm{P}$. Next, we need to determine the radial velocity of the gas at $\mathrm{P}$ relative to the Sun by projecting the velocity vector, $\vec{v}$, on to the line-of-sight vector, $\vec{d}$, and subtracting the projected contribution from the Sun's own orbital velocity, thus
\begin{equation}
	v_{\rm rad} = \frac{\vec{d} \scalprod (\vec{v} - \vec{v}_{\sun})}{|\vec{d}|} = \frac{(\vec{r} - \vec{r}_{\sun}) \scalprod (\vec{v} - \vec{v}_{\sun})}{d} \, . \label{eqn:vrad-def}
\end{equation}
Inserting Eq.~\ref{eqn:radius} into Eq.~\ref{eqn:vrad-def}, we obtain the following expression for the radial velocity:
\begin{equation}
	v_{\rm rad} = \left[ v_{\rm rot}(r_{xy}) \frac{r_{\sun}}{r_{xy}} - v_{\sun} \right] \sin(l) \cos(b) \label{eqn:vrad}
\end{equation}
where $v_{\rm rot}(r_{xy})$ is the rotation curve of the Milky Way and $r_{xy} = |(r_{x},r_{y})|$. Note that the geometry used here implies cylindrical rotation, i.e.\ the rotation velocity of the disc is assumed to be independent of height, $z$, above the Galactic plane.

For this work, the rotation curve of \citet{Clemens1985} is used, specifically the polynomial fit to the rotation curve for the IAU standard values of $r_{\sun} = 8.5~\mathrm{kpc}$ and $v_{\sun} = 220~\mathrm{km \, s}^{-1}$. The rotation curve of \citet{Clemens1985} was derived from combined CO and \ion{H}{I} observations of the Milky Way and should therefore be well-suited to describe the rotation velocity of the neutral gas disc for the purpose of identifying high-velocity emission.

With the radial velocity of any arbitrary point at hand, we can now determine the velocity range of Galactic disc gas for any position in the sky by simply moving away from the Sun in discrete steps along the line of sight, determining the radial velocity of the gas in each step, and recording the maximum and minimum values, $v_{\rm max}$ and $v_{\rm min}$, encountered in between the Sun and the boundary of the cylindrical disc model. Lastly, the derived velocity range will need to be expanded by a fixed deviation velocity, $v_{\rm dev}$, to account for the intrinsic velocity dispersion of the disc gas as well as any small deviations from the regular rotation velocity of the disc \citep{Wakker1991a}. The final velocity range, $[v_{\rm min} - v_{\rm dev}, v_{\rm max} + v_{\rm dev}]$, can then be masked in the data cube, leaving only channels with velocities inconsistent with Galactic rotation.

\begin{figure*}
	\centering
	\includegraphics[width=\linewidth]{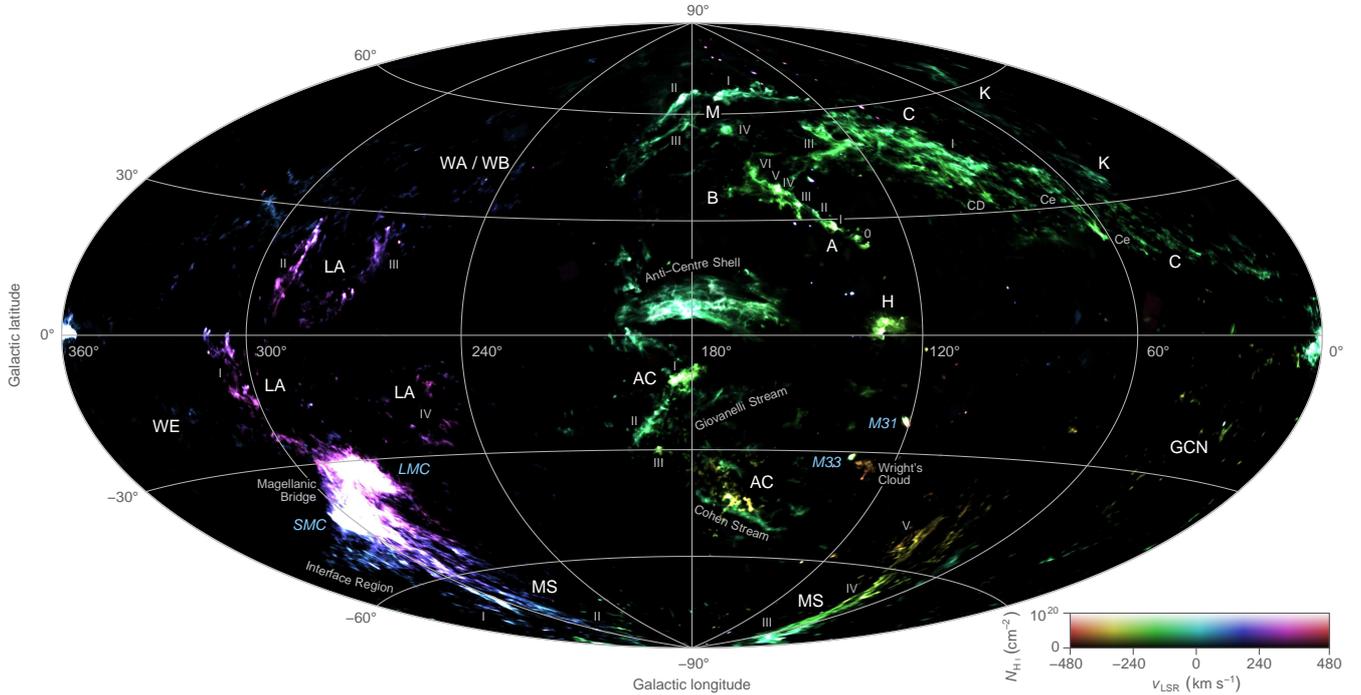}
	\caption{All-sky false-colour map of high-velocity gas presented in Hammer--Aitoff projection in Galactic coordinates centred on the Galactic anti-centre. Brightness and hue in the image represent \ion{H}{I} column density (linear from $0$ to $10^{20}~\mathrm{cm}^{-2}$) and LSR radial velocity of the emission, respectively. Several major HVC complexes as well as a few notable individual structures and external galaxies are labelled.}
	\label{fig:colour-map}
\end{figure*}

\subsection{Optimisation of model parameters}

While we now have a simple model of the Galactic disc that will allow us to mask and remove Galactic \ion{H}{I} emission, there are several free parameters, specifically the disc size, $r_{\rm max}$ and $z_{\rm max}$, as well as the value of the deviation velocity, $v_{\rm dev}$, that will need to be chosen first. For this purpose, a low-resolution copy of the entire HI4PI survey was created by binning the data by a factor of~8 in the spatial domain and~4 in the spectral domain. This reduced the total data volume by a factor of 256, allowing for fast processing of the full sky at low resolution.

Next, maps of HVC emission were calculated from the binned data using a matrix of different values for the three free parameters. The following set of parameter values were tested in this way: $r_{\rm max} = 20$, $25$ and $30~\mathrm{kpc}$, $z_{\rm max} = 2$, $5$ and $10~\mathrm{kpc}$, and $v_{\rm dev} = 50$, $60$ and $70~\mathrm{km \, s}^{-1}$. This resulted in a total of 27~different parameter combinations to be processed and tested. The HVC maps resulting from the different combinations of parameters were finally inspected by eye to determine the parameter values that would successfully remove most of the Galactic foreground \ion{H}{I} emission while retaining as much of the \ion{H}{I} emission from the HVC population as possible.

The resulting optimal set of disc parameters is $r_{\rm max} = 20~\mathrm{kpc}$, $z_{\rm max} = 5~\mathrm{kpc}$ and $v_{\rm dev} = 70~\mathrm{km \, s}^{-1}$. Interestingly, varying the value of $r_{\rm max}$ did not have any significant influence on the result, while changing the thickness of the disc led to substantial changes in the purity of the resulting HVC map. Similarly, choosing a lower deviation velocity would introduce significant residuals from intermediate-velocity gas, e.g.\ in the region of complex~M. Compared to the parameters adopted by \citet{Wakker1991a}, the disc radius chosen here is smaller, while the thickness of the disc and deviation velocity are larger. This results in a `cleaner' map with lower levels of residual emission from the Galactic disc (e.g.\ in the region of the Outer Arm), while the major HVC complexes are largely unaffected due to their significant kinematic separation from the disc.

\subsection{Generation of HVC maps}

After determination of the optimal set of parameters for the Galactic disc model, a custom script written in C++ was used to process the original HI4PI data with the aim of generating an all-sky map of HVCs. The script first reads an HI4PI sub-cube of $20\degr \times 20\degr$ into memory. It then loops across all spatial pixels of the cube, extracts the Galactic coordinates at each pixel, and applies Eq.~\ref{eqn:vrad} to determine the velocity range of Galactic emission at that position by moving outwards from the Sun along the line of sight in steps of $100~\mathrm{pc}$. All spectral channels located within that Galactic velocity range, expanded by the deviation velocity of $70~\mathrm{km \, s}^{-1}$, are then masked before the processed cube is written back to disk for subsequent analysis.

In order to facilitate the creation of two-dimensional maps of the HVC sky from the masked data cubes, the masked cubes were first smoothed spatially by a Gaussian of $48~\mathrm{arcmin}$ FWHM and spectrally by a boxcar filter 5~channels (approximately $7.5~\mathrm{km \, s}^{-1}$) wide. Only those pixels of the original cube that had $T_{\rm B} > 50~\mathrm{mK}$ in the smoothed copy of the cube were used in the generation of any two-dimensional maps created from the data cube. This approach ensures that the faint outer parts of diffuse \ion{H}{I} sources are included in the map, while compact artefacts such as noise and radio frequency interference (RFI) are largely excluded. Lastly, any two-dimensional HVC maps created from the masked data cubes were concatenated to form a single, all-sky map.

The resulting two-dimensional all-sky HVC map in Hammer--Aitoff projection is presented in Fig.~\ref{fig:colour-map} in Galactic coordinates centred on the Galactic anti-centre. In this colour map, which was created to facilitate visual inspection of the HVC emission, brightness represents \ion{H}{I} column density, while hue reflects the LSR radial velocity of the emission. This allows HVCs of different velocities to be distinguished by their different hues in regions of spatially overlapping emission. The naming of HVC features in the map is in accordance with \citet{Wakker2001}, \citet{Putman2003}, \citet{Wakker2004}, \citet{Bruens2005}, \citet{Venzmer2012} and \citet{For2013}.

To provide data products that are more suitable for a quantitative analysis, a two-dimensional map of the $0^{\rm th}$ spectral moment was created as well. The $0^{\rm th}$ moment map was converted to \ion{H}{I} column density in units of $\mathrm{cm}^{-2}$, using the standard conversion factor of $1.823 \times 10^{18} \, \mathrm{cm}^{-2} / (\mathrm{K \, km \, s}^{-1})$, under the assumption that the emission is optically thin and fills the $16.2~\mathrm{arcmin}$ beam. A labelled all-sky view of the resulting column density map in Hammer--Aitoff projection in Galactic coordinates is presented in the upper panel of Fig.~\ref{fig:all-sky-maps} in Appendix~\ref{app:maps}.

\begin{figure*}
	\centering
	\includegraphics[width=\linewidth]{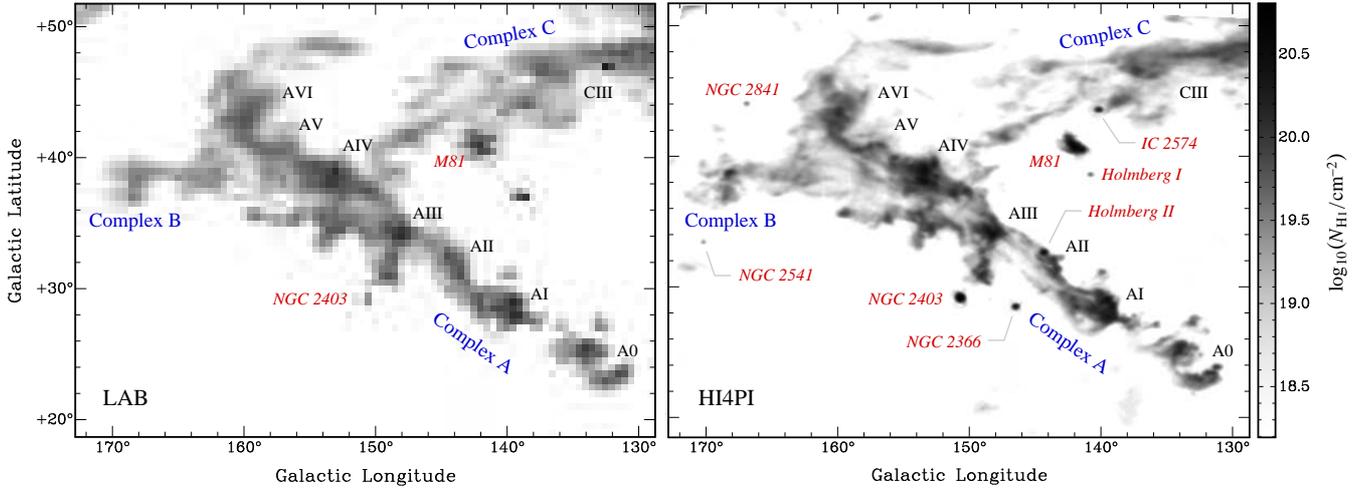}
	\caption{Comparison between the old HVC map from \citet{Westmeier2007} based on LAB data (left) and the new map presented here based on HI4PI data (right) in the region of HVC complex~A. The substantial improvement in angular resolution reveals a complex network of intricate gas filaments and clumps that are not resolved by the old LAB data. Note the large number of nearby galaxies visible in the high-resolution HI4PI data, the brightest of which have been labelled.}
	\label{fig:comparison}
\end{figure*}

In a similar fashion one could generate a radial velocity map of the HVC sky from the $1^{\rm st}$ spectral moment. However, a general issue with the $1^{\rm st}$ moment is that the resulting values may be highly susceptible to the influence of noise in cases where either the \ion{H}{I} signal is faint or the applied mask is too large. In addition, the $1^{\rm st}$ moment may produce arbitrary values in cases where multiple sources of emission exist along the line of sight. In order to avoid these issues, a velocity map was created by fitting a single Gaussian function to the brightest line component in each spectrum using a custom script written in \software{Python}. The script will first identify the location of the brightest emission in the spectrum, then determine the mean and standard deviation of the emission within a window of $\pm 20$~channels around that position, and finally use those initial estimates to fit a Gaussian function to the entire spectrum. Successful fits are accepted for spectra that have positive integrated flux, a peak brightness temperature exceeding three times the mean RMS of the HI4PI survey (i.e.\ $T_{\rm B} \ge 130~\mathrm{mK}$), and an initial estimate of the standard deviation of at least two channels (i.e.\ $\sigma \gtrsim 2.6~\mathrm{km \, s}^{-1}$). The amplitude, mean and standard deviation from accepted Gaussian fits are then written into a FITS file for further analysis.

As the native LSR velocity frame of the HI4PI data is dominated by the projected rotation velocity of the Galactic disc, the resulting velocity map was converted to the more appropriate Galactic standard-of-rest (GSR) frame in units of $\mathrm{km \, s}^{-1}$, assuming the standard orbital velocity of $v_{\sun} = 220~\mathrm{km \, s}^{-1}$ in the direction of $(l, b) = (90^{\circ}, 0^{\circ})$. The resulting GSR velocity map of the HVC sky in Galactic coordinates is presented in the bottom panel of Fig.~\ref{fig:all-sky-maps} in Appendix~\ref{app:maps}.

Finally, in order to facilitate studies of the Magellanic system, the \ion{H}{I} column density and GSR velocity maps were converted from the native Galactic coordinate system, $(l, b)$, to Magellanic coordinates, $(l_{\rm Mag}, b_{\rm Mag})$, as defined by \citet{Nidever2008}. The resulting maps are presented in Fig.~\ref{fig:all-sky-maps-2} in Appendix~\ref{app:maps}. The Magellanic coordinate system is defined such that the Magellanic Stream and Leading Arm are approximately aligned with the equator at $b_{\rm Mag} \approx 0\degr$, while the Large Magellanic Cloud defines the origin of the longitudinal axis at $l_{\rm Mag} = 0\degr$. This avoids the usually strong distortions of parts of the Magellanic system in standard equatorial and Galactic coordinates and provides a more natural coordinate system suitable for studies of the Magellanic system. Note that the maps in Fig.~\ref{fig:all-sky-maps-2} have been centred on $l_{\rm Mag} = 270\degr$ instead of $0\degr$ to prevent Complex~C from being wrapped around the edge of the map.

\begin{figure*}
	\centering
	\includegraphics[width=0.88\linewidth]{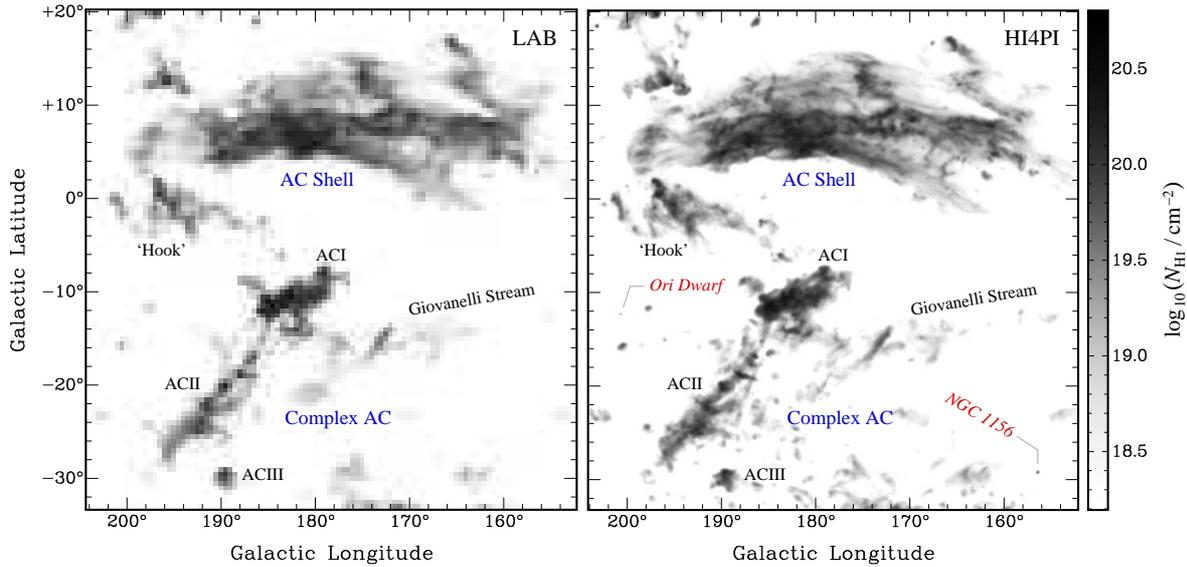}
	\caption{Same as in Fig.~\ref{fig:comparison}, but for a larger region around the northern section of the anti-centre (AC) complex and the adjacent anti-centre shell.}
	\label{fig:comparison2}
\end{figure*}

\section{Results and discussion}
\label{sect_discussion}

\subsection{The HVC sky}

The all-sky maps shown in Fig.~\ref{fig:colour-map} as well as Fig.~\ref{fig:all-sky-maps} and~\ref{fig:all-sky-maps-2} show the major HVC complexes in the sky in much greater detail than ever before. Dominant structures in the maps include the large HVC complexes~A, C and~M in the northern (celestial and Galactic) hemisphere, the anti-centre complex and shell near the Galactic anti-centre, and the Magellanic Clouds with the extended structures of the Magellanic Stream and Leading Arm mostly confined to the southern hemisphere. A few smaller or less sharply defined structures are visible as well, including HVC complexes~H and~K. Several HVC complexes with low deviation velocities, most of which kinematically overlap with Galactic disc emission, are largely missing from the map, including complexes~L, WA, WB, WC, G, R, GCP as well as parts of complex~H and the Giovanelli Stream.

The final maps are almost entirely free of residual emission from the Galactic disc. Notable contamination stems from peculiar-velocity gas near the Galactic centre, faint intermediate-velocity gas filaments near the north Galactic pole, and a few remnants of extra-planar gas associated with the Outer Arm near $(l, b) = (80^{\circ}, 25^{\circ})$. Moreover, not all emission seen in the maps is due to high-velocity \ion{H}{I} gas. On the one hand, low levels of RFI and residual stray radiation may still be present in some parts of the sky, in particular in the northern celestial hemisphere covered by EBHIS. On the other hand, several nearby galaxies at low radial velocities are included in the map as well, most prominently the Large and Small Magellanic Cloud as well as M31 and M33.

It should be noted that some of the HVC complexes partially extend into the velocity range of Galactic disc gas, which inevitably leads to the loss of some emission from the maps. As a consequence, the column density values in some regions of Fig.~\ref{fig:all-sky-maps} and~\ref{fig:all-sky-maps-2} may be lower limits while the velocities may be inaccurate. The Magellanic Stream in particular crosses from negative to positive LSR velocities near the south Galactic pole, inevitably resulting in flux loss. For the same reason, some neutral gas clouds located in the Galactic halo may be missing entirely, as their kinematics in combination with the viewing geometry may have placed them within the deviation velocity range applied here. This would effectively `hide' those clouds within the velocity range of the Galactic disc, where they would no longer be discernible as high-velocity gas.

\subsection{Comparison with LAB data}

Using the regions around HVC complex~A and the anti-centre (AC) complex as an example, Fig.~\ref{fig:comparison} and~\ref{fig:comparison2} compare the quality of the previous all-sky map created by \citet{Westmeier2007} based on LAB data with the new HI4PI-based map presented here. While the individual clouds of complex~A were barely resolved by the $36~\mathrm{arcmin}$ beam of the LAB survey, the new data resolve complex~A into a remarkable network of intricate \ion{H}{I} filaments and small clumps. Assuming a distance in the range of $4$ to $10~\mathrm{kpc}$ for complex~A \citep{vanWoerden1999}, the corresponding physical resolution would be in the range of about $20$ to $50~\mathrm{pc}$. A similar effect can be seen in the anti-centre region, where the AC shell in particular breaks up into a system of fine, interwoven gas filaments that were too narrow to be resolved by the $36~\mathrm{arcmin}$ beam of the LAB survey in the past.

An additional improvement in the quality of the map can be attributed to the full spatial sampling of the HI4PI survey data in contrast to the LAB data which were only sampled on a $30~\mathrm{arcmin}$ grid. The undersampling of the LAB survey is clearly visible in the left-hand panels of Fig.~\ref{fig:comparison} and~\ref{fig:comparison2}. On top of the improvements in resolution and sampling, the HI4PI survey is also more sensitive than the LAB survey, resulting in a better column density sensitivity\footnote{\ion{H}{I} column density sensitivities are specified for a signal of $5 \times \sigma_{\rm RMS}$ across $20~\mathrm{km \, s}^{-1}$ under the assumption that the emission is optically thin and fills the beam size.} of about $2.3 \times 10^{18}~\mathrm{cm}^{-2}$ (as compared to about $3.7 \times 10^{18}~\mathrm{cm}^{-2}$ for the LAB survey across a much larger beam solid angle). The significant improvement in both angular resolution and sensitivity is also highlighted by the large number of nearby galaxies visible in the HI4PI map in the right-hand panel of Fig.~\ref{fig:comparison}, only the brightest and most extended of which are also seen in the LAB data.

Lastly, the general agreement between the LAB and HI4PI maps is remarkable in view of the fact that a slightly different methodology was used by \citet{Westmeier2007} for generating the LAB-based HVC map. Nevertheless, the same general features are visible in the two maps, indicating that most of the HVC structures in the sky are kinematically well-separated from the Galactic disc and thus appear in both maps.

\begin{figure*}
	\centering
	\includegraphics[width=0.49\linewidth]{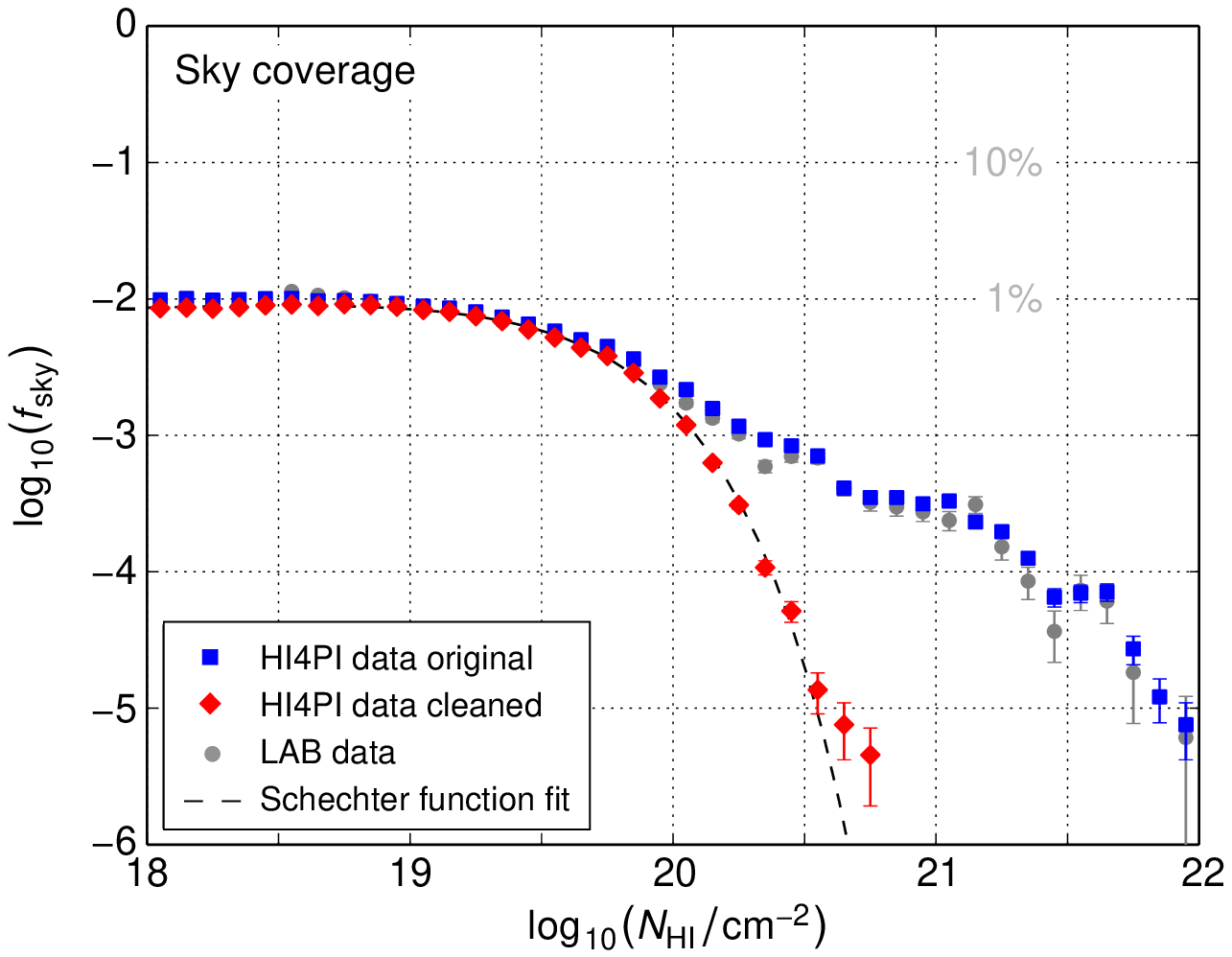}
	\includegraphics[width=0.49\linewidth]{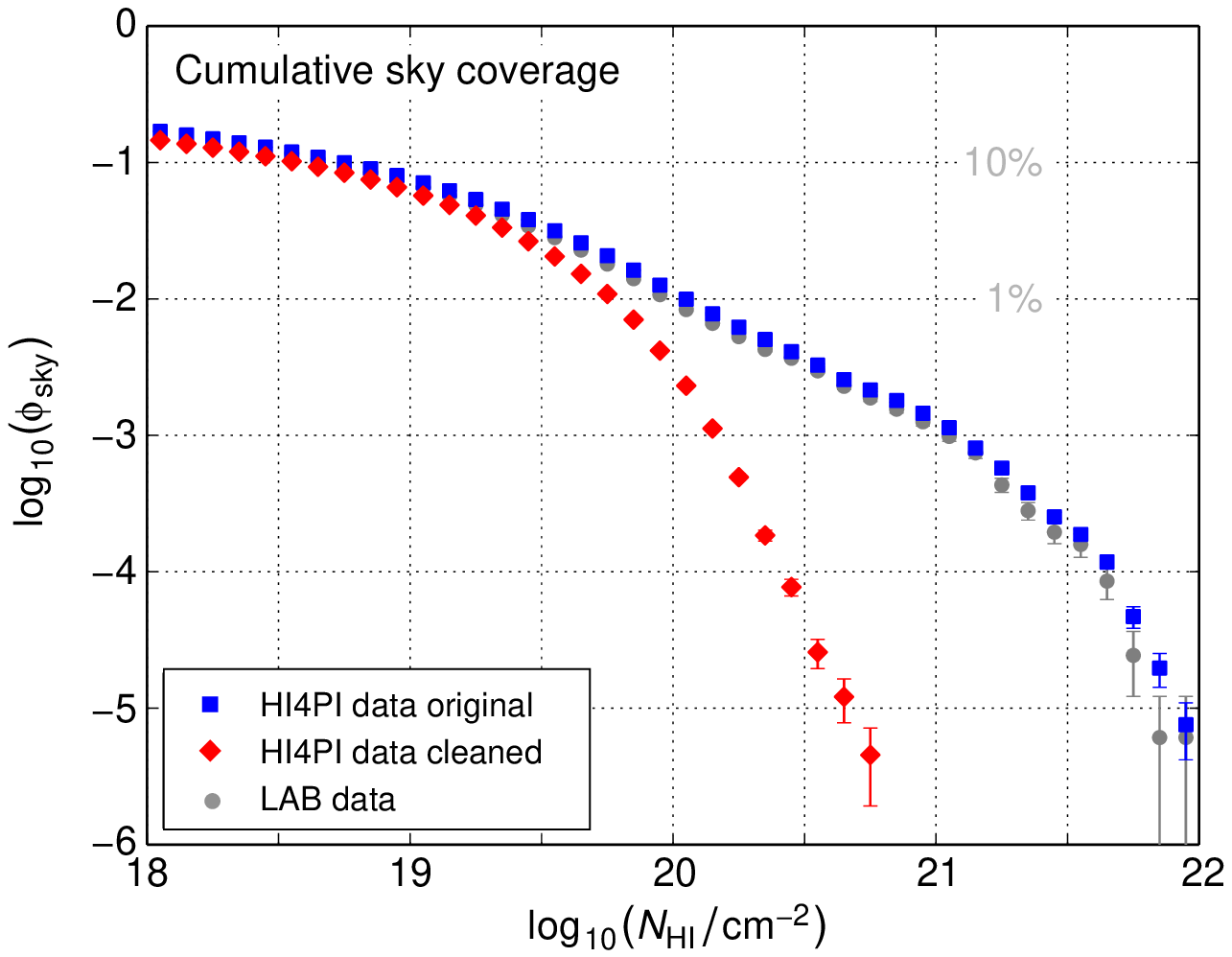}
	\caption{\emph{Left:} Sky coverage fraction, $f_{\rm sky}$, of high-velocity gas as a function of \ion{H}{I} column density in bins of $\Delta \log_{10}(N_{\ion{H}{I}} / \mathrm{cm}^{-2}) = 0.1$. \emph{Right:} Cumulative sky coverage fraction, $\phi_{\rm sky}$, of high-velocity gas exceeding a given column density level. In both panels the blue squares represent the original data, while the red diamonds were derived from a data set with bright galaxies (including the Magellanic Clouds), stray radiation artefacts and residual Galactic emission from the Outer Arm and the Galactic centre region removed. The LAB-based HVC map from \citet{Westmeier2007} is shown as the grey circles for comparison. Error bars are indicative of Poisson errors. The dashed, black line in the left-hand panel shows a Schechter function fitted to the red diamonds.}
	\label{fig:sky-coverage}
\end{figure*}

\subsection{Sky coverage fraction of HVCs}

From the resulting HVC map one can immediately derive the sky coverage fraction of high-velocity \ion{H}{I} emission as a function of column density. For this purpose, column density values were extracted from the map in \emph{plate carr\'{e}e} projection on a grid of $\Delta b = 0.25~\mathrm{deg}$ in Galactic latitude and $\Delta l \cos(b) = 0.25~\mathrm{deg}$ in Galactic longitude, resulting in a total of about $4 \uppi / \Omega_{\rm grid} \approx 660,000$ largely independent data points across the full sky. The resulting sky coverage fraction, $f_{\rm sky}$, and cumulative sky coverage fraction, $\phi_{\rm sky}$, of HVC emission in excess of a given \ion{H}{I} column density level are shown as the blue squares in Fig.~\ref{fig:sky-coverage}. The cumulative sky coverage fraction for HVC emission of $N_{\ion{H}{I}} > 2 \times 10^{18}~\mathrm{cm}^{-2}$ is approximately 15~per cent. It decreases to about 7~per cent at $N_{\ion{H}{I}} > 10^{19}~\mathrm{cm}^{-2}$ and about 1~per cent at $N_{\ion{H}{I}} > 10^{20}~\mathrm{cm}^{-2}$.

The original HVC map still contains some emission from sources other than HVCs, in particular nearby galaxies, residual emission from the Galactic disc and stray radiation residuals in the northern celestial hemisphere covered by EBHIS. Therefore, a cleaned HVC map was created by manually masking bright galaxies (including the LMC, SMC and Magellanic Bridge, but not the Interface Region, Magellanic Stream and Leading Arm), bright emission near the Galactic centre, residual emission from the Outer Arm and bright stray radiation residuals in the EBHIS data. The measurements of $f_{\rm sky}$ and $\phi_{\rm sky}$ resulting from the cleaned map are shown as the red diamonds in Fig.~\ref{fig:sky-coverage}. Most of the high-column density emission above $10^{20}~\mathrm{cm}^{-2}$ in the original map can be attributed to the Magellanic Clouds and Bridge, and there is no more emission above about $5 \times 10^{20}~\mathrm{cm}^{-2}$ in the cleaned map. At the same time, removing known non-HVC emission from the map did not greatly change the column density distribution below about $3 \times 10^{19}~\mathrm{cm}^{-2}$. The resulting coverage fractions from the cleaned map are $\phi_{\rm sky} \approx 13$, 6 and 0.2~per cent for column densities of $N_{\ion{H}{I}} > 2 \times 10^{18}$, $10^{19}$ and $10^{20}~\mathrm{cm}^{-2}$, respectively.

These results are comparable to the cumulative sky coverage of 15~per cent at $N_{\ion{H}{I}} > 2 \times 10^{18}~\mathrm{cm}^{-2}$ reported by \citet{Wakker2004}. Similar numbers were found by \citet{Wakker1991a} who derived either 18 or 11~per cent above $T_{\rm B} = 50~\mathrm{mK}$ (equivalent to about $2 \times 10^{18}~\mathrm{cm}^{-2}$) depending on whether the Magellanic Stream and Outer Arm were included or excluded from the analysis. A larger sky coverage fraction of 37~per cent above a significantly lower $5 \, \sigma$ column density threshold of $7 \times 10^{17}~\mathrm{cm}^{-2}$ was found by \citet{Murphy1995} from pointed \ion{H}{I} observations towards 102~quasars (also see \citealp{Lockman2002}).

For comparison, the analysis was repeated on the all-sky HVC map from \citet{Westmeier2007} based on the LAB survey. The results are presented as the grey circles in Fig.~\ref{fig:sky-coverage} for the full map with the Magellanic Clouds included. The similarity between the HI4PI and LAB sky coverage factors is remarkable, confirming that both maps essentially trace the same structures on the sky. Upon closer inspection the LAB data points generally appear to be slightly below the HI4PI data points. The most likely explanation for this discrepancy is that the $36~\mathrm{arcmin}$ beam size of the LAB survey is significantly larger than the $16.2~\mathrm{arcmin}$ HI4PI beam, resulting in beam smearing effects and thus generally lower column density levels in the LAB survey for any emission that is not diffuse across the beam. Therefore, the LAB sky coverage curves in Fig.~\ref{fig:sky-coverage} are actually expected to be shifted slightly to the left in the direction of lower column densities.

Looking at the differential sky coverage fraction after removal of galaxies and artefacts again (red diamonds in the left-hand panel of Fig.~\ref{fig:sky-coverage}), it would appear that the sky coverage of HVCs essentially follows a power law at lower column densities, with a distinct turnover point at just under $10^{20}~\mathrm{cm}^{-2}$. We can therefore fit a Schechter function \citep{Schechter1976} of the form
\begin{equation}
	f(n) = \ln(10) f^{\star} n^{\alpha + 1} \exp(-n)
\end{equation}
to the data points in $\log_{10}(N_{\ion{H}{I}} / \mathrm{cm}^{-2})$ space to extract the basic parameters of this truncated power law. Here, $n = N_{\ion{H}{I}} / N_{\ion{H}{I}}^{\star}$ is the dimensionless \ion{H}{I} column density, $N_{\ion{H}{I}}^{\star}$ denotes the location of the characteristic turnover point, $\alpha$ is the exponent of the power-law component of the function, and $f^{\star}$ is a global normalisation factor.

The fitted Schechter function is shown as the dashed, black line in the left-hand panel of Fig.~\ref{fig:sky-coverage} and yields a power-law exponent of $\alpha = -0.93$, a characteristic turnover point of $\log_{10}(N_{\ion{H}{I}}^{\star} / \mathrm{cm}^{-2}) = 19.7$ and a scaling factor of $f^{\star} = 5.0 \times 10^{-3}$ normalised per interval of $\Delta \log_{10}(N_{\ion{H}{I}} / \mathrm{cm}^{-2}) = 0.1$. Hence, at lower column densities, the differential sky coverage fraction is almost flat in logarithmic space (i.e., the power-law exponent is close to $-1$), with a turnover point at a characteristic column density level of $N_{\ion{H}{I}}^{\star} \approx 5 \times 10^{19}~\mathrm{cm}^{-2}$. Consequently, there is no \emph{preferred} column density of HVC emission. There is also no indication of another turnover near the sensitivity limit of the HI4PI data ($\approx 10^{18}~\mathrm{cm}^{-2}$), which is consistent with the findings of \citet{Lockman2002} who, at a slightly lower median completeness limit of $8 \times 10^{17}~\mathrm{cm}^{-2}$, report HVC emission in 37~per cent of their 860 quasar sightlines.

\begin{table*}
	\caption{List of FITS data products made available as supplementary material in the online version of this paper. The standard FITS projection codes used are \texttt{CAR} (\emph{plate carr{\'e}e}) and \texttt{AIT} (Hammer--Aitoff). File sizes are given for the compressed files in units of $1~\mathrm{MB} = 2^{20}~\mathrm{B}$.}
	\centering
	\begin{tabular}{cccccc}
		\hline
		Data product & Coordinate & Proj.          & Dimension          & File name                               & File size \\
		             & system     &                & (pixels)           &                                         & (MB)      \\
		\hline
		\lognhi      & Galactic   & \texttt{CAR}   & $4323 \times 2144$ & \texttt{hi4pi-hvc-nhi-gal-car.fits.gz}  & 7.7 \\
		\lognhi      & Galactic   & \texttt{AIT}   & $3891 \times 1947$ & \texttt{hi4pi-hvc-nhi-gal-ait.fits.gz}  & 4.6 \\
		\lognhi      & Magellanic & \texttt{CAR}   & $4323 \times 2144$ & \texttt{hi4pi-hvc-nhi-mag-car.fits.gz}  & 7.9 \\
		\vel{LSR}    & Galactic   & \texttt{CAR}   & $4323 \times 2144$ & \texttt{hi4pi-hvc-vlsr-gal-car.fits.gz} & 4.5 \\
		\vel{LSR}    & Galactic   & \texttt{AIT}   & $3891 \times 1947$ & \texttt{hi4pi-hvc-vlsr-gal-ait.fits.gz} & 2.9 \\
		\vel{GSR}    & Galactic   & \texttt{CAR}   & $4323 \times 2144$ & \texttt{hi4pi-hvc-vgsr-gal-car.fits.gz} & 4.6 \\
		\vel{GSR}    & Galactic   & \texttt{AIT}   & $3891 \times 1947$ & \texttt{hi4pi-hvc-vgsr-gal-ait.fits.gz} & 3.0 \\
		\vel{GSR}    & Magellanic & \texttt{CAR}   & $4323 \times 2144$ & \texttt{hi4pi-hvc-vgsr-mag-car.fits.gz} & 4.7 \\
		\hline
	\end{tabular}
	\label{tab:data-products}
\end{table*}

\section{Data release}
\label{sect_release}

All-sky HVC maps are a useful tool for a wide range of scientific applications, including, among others, the study of the HVC population itself \citep{Wakker1991a,Wakker1991b}, comparison of the \ion{H}{I} emission with optical and ultra-violet absorption lines of ionised gas in the Galactic halo \citep{Fox2006,Lehner2012}, cross-correlation of the \ion{H}{I} emission with X-ray emission \citep{Kerp1996,Kerp1999,Shelton2012} and H$\upalpha$ emission \citep{Tufte1998,Barger2012}, and comparison of the \ion{H}{I} gas with molecular gas \citep{Wakker1997b} and dust emission \citep{Wakker1986,MivilleDeschenes2005,Williams2012,Lenz2016} in the quest for evidence of star formation in HVC complexes such as the Magellanic Stream \citep{Tanaka1982,Brueck1983} and Leading Arm \citep{Casetti-Dinescu2014}.

In order to facilitate a wider scientific use of the new all-sky HVC maps by the community, \ion{H}{I} column density maps and radial velocity maps in the LSR and GSR frames have been made publicly available in both \emph{plate carr\'{e}e} and Hammer--Aitoff projections (see Table~\ref{tab:data-products} for details). The maps are provided as two-dimensional FITS images (\emph{Flexible Image Transport System}; \citealp{Wells1981}) and are available as supplementary material in the online version of this paper. Column density and GSR velocity maps in \emph{plate carr\'{e}e} projection are also made available in Magellanic coordinates (using the definition of \citealp{Nidever2008}) in which the Magellanic Stream is aligned with the equator to facilitate studies of the Magellanic system. Note that for convenience the corresponding FITS files have Galactic coordinates defined in their header, as the FITS standard does not natively support the Magellanic coordinate system \citep{Calabretta2002}, while user-specific coordinate systems may not be supported by some of the third-party software commonly used to handle and display FITS images.

\section{Summary}
\label{sect_summary}

In this paper a new all-sky map of Galactic HVCs based on the HI4PI survey is presented. For this purpose, a simple, cylindrical model of the Galactic disc with a disc radius of $20~\mathrm{kpc}$ and a disc height of $5~\mathrm{kpc}$ was created based on which the expected velocity range of Galactic emission was determined and masked after applying an additional deviation velocity of $70~\mathrm{km \, s}^{-1}$. All-sky \ion{H}{I} column density and velocity maps of high-velocity gas were then generated from the masked cubes.

The resulting all-sky maps show the Milky Way's HVC population at an unprecedented angular resolution of $16.2~\mathrm{arcmin}$ and a $5 \sigma$ column density sensitivity of about $2.3 \times 10^{18}~\mathrm{cm}^{-2}$. Most of the HVC complexes are resolved into a network of \ion{H}{I} filaments and clumps not seen in the previous HVC maps of \citet{Wakker1991a} and \citet{Westmeier2007}. The overall sky coverage fraction of high-velocity gas is approximately 15~per cent for emission of $N_{\ion{H}{I}} > 2 \times 10^{18}~\mathrm{cm}^{-2}$, decreasing to about 13~per cent when the Magellanic Clouds and some additional non-HVC emission are excluded. The differential HVC sky coverage fraction as a function of column density essentially follows a truncated power law with an exponent of $-0.93$ and a characteristic turnover point at a column density level of about $5 \times 10^{19}~\mathrm{cm}^{-2}$.

FITS files of \ion{H}{I} column density and velocity maps of the HVC sky in different coordinate systems and projections have been made publicly available to facilitate the scientific use of the new HVC map by the entire community. Users are requested to include a reference to this paper in any publication making use of these data files.

\section*{Acknowledgements}

The author would like to thank L.~Staveley-Smith for valuable discussions and comments on the manuscript. This work is based on publicly released data from the HI4PI survey which combines the Effelsberg--Bonn \ion{H}{I} Survey (EBHIS) in the northern hemisphere with the Galactic All-Sky Survey (GASS) in the southern hemisphere. This publication is based on observations with the 100-m telescope of the MPIfR (Max-Planck-Institut f\"{u}r Radioastronomie) at Effelsberg. The Parkes Radio Telescope is part of the Australia Telescope which is funded by the Commonwealth of Australia for operation as a National Facility managed by CSIRO. This research has made use of the VizieR catalogue access tool, CDS, Strasbourg, France. The original description of the VizieR service was published in A\&AS~143, 23. This research has made use of NASA's Astrophysics Data System Bibliographic Services. This research has made use of the NASA/IPAC Extragalactic Database (NED), which is operated by the Jet Propulsion Laboratory, California Institute of Technology, under contract with the National Aeronautics and Space Administration.




\bibliographystyle{mnras}
\bibliography{westmeier_hvc-sky} 




\section*{Supporting information}

Additional Supporting Information may be found in the online version of this paper:

As detailed in Section~\ref{sect_release}, FITS images of the new all-sky map of HVCs presented in this paper are made available as supplementary material in the online version of this paper. An overview of the individual data products and corresponding FITS files is provided in Table~\ref{tab:data-products}.

Please note: Oxford University Press are not responsible for the content or functionality of any supporting materials supplied by the authors. Any queries (other than missing material) should be directed to the corresponding author for the paper.

\appendix

\section{Column density and velocity maps}
\label{app:maps}

\ion{H}{I} column density and GSR radial velocity maps of the high-velocity sky in Hammer--Aitoff projection in both Galactic and Magellanic coordinates are presented in this appendix.

\begin{figure*}
	\centering
	\includegraphics[width=\linewidth]{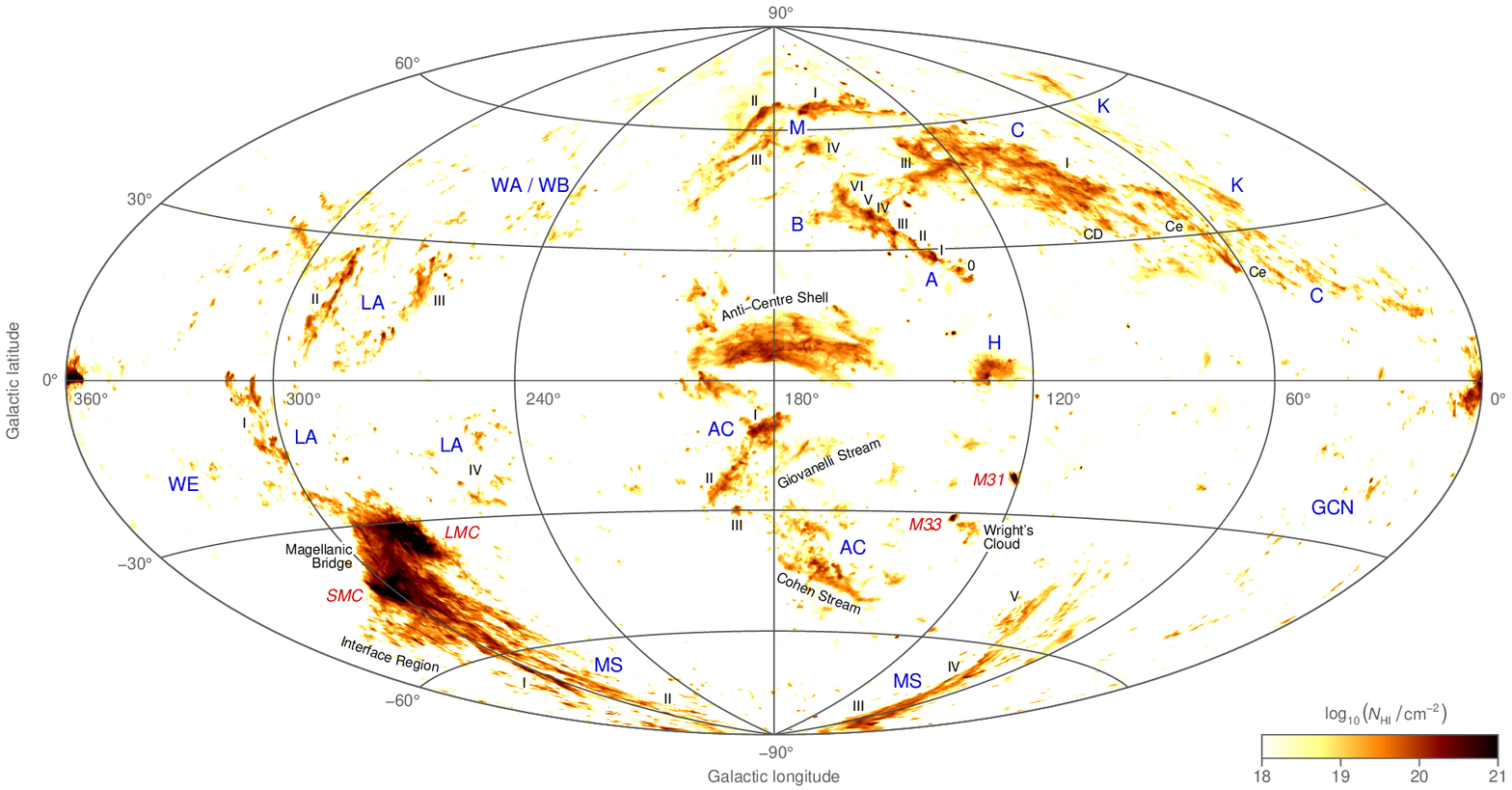}
	\includegraphics[width=\linewidth]{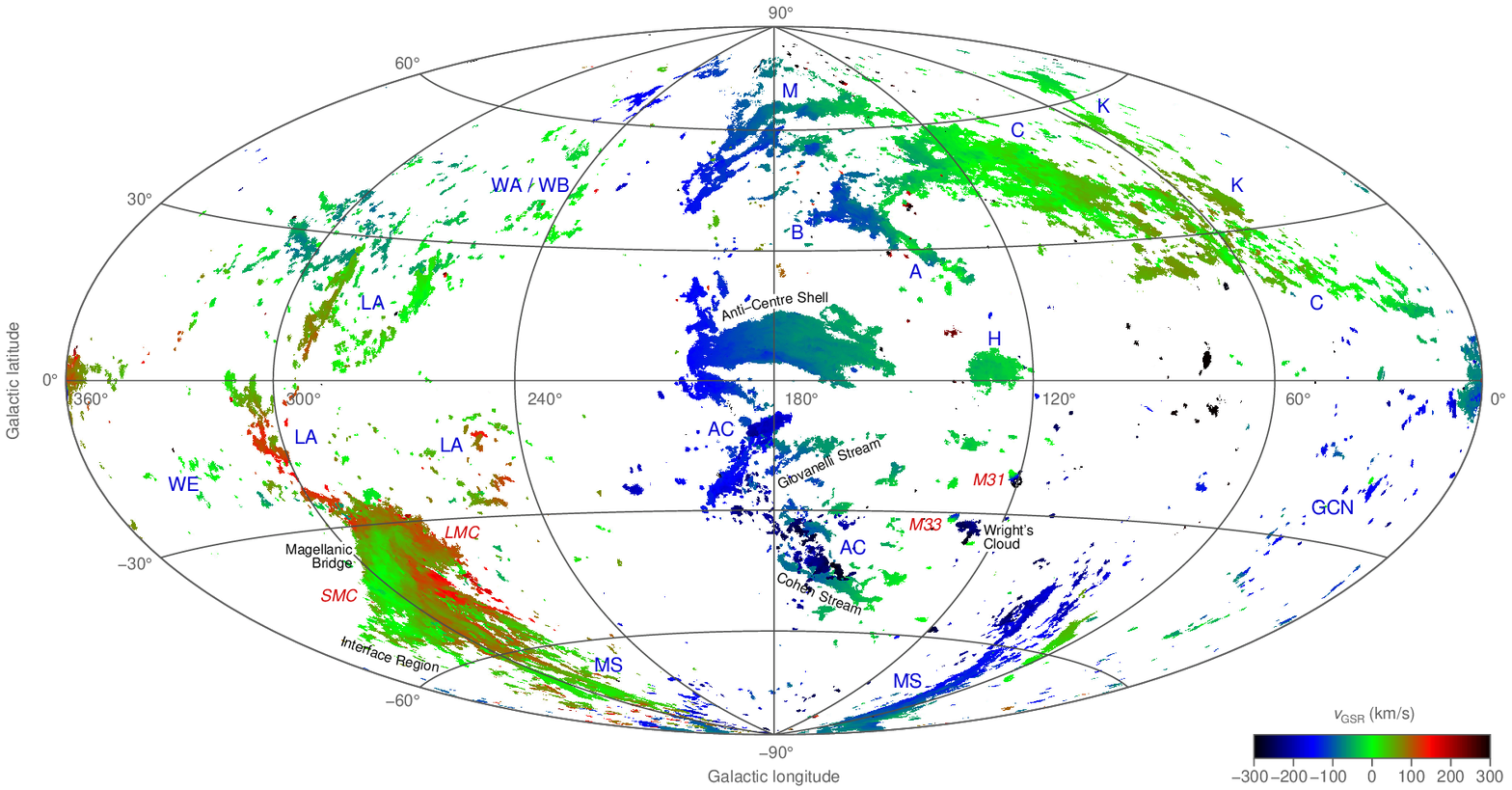}
	\caption{All-sky maps of high-velocity gas presented in Hammer--Aitoff projection in Galactic coordinates centred on the Galactic anti-centre. Top: \ion{H}{I} column density, derived from the $0^{\mathrm{th}}$ spectral moment, in the range of $\log_{10}(N_{\ion{H}{I}} / \mathrm{cm}^{-2}) = 18$ to $21$ under the assumption that the emission is optically thin and fills the $16.2$-arcmin beam. Bottom: Radial velocity in the GSR frame, derived from Gaussian fits, in the range of $v_{\rm GSR} = -300$ to $+300~\mathrm{km \, s}^{-1}$. Several major HVC complexes as well as a few notable individual structures and external galaxies are labelled. Note that a few remaining artefacts caused by RFI and residual stray radiation were manually removed from both maps for presentation purposes, but they are still present in the released FITS images.}
	\label{fig:all-sky-maps}
\end{figure*}

\begin{figure*}
	\centering
	\includegraphics[width=\linewidth]{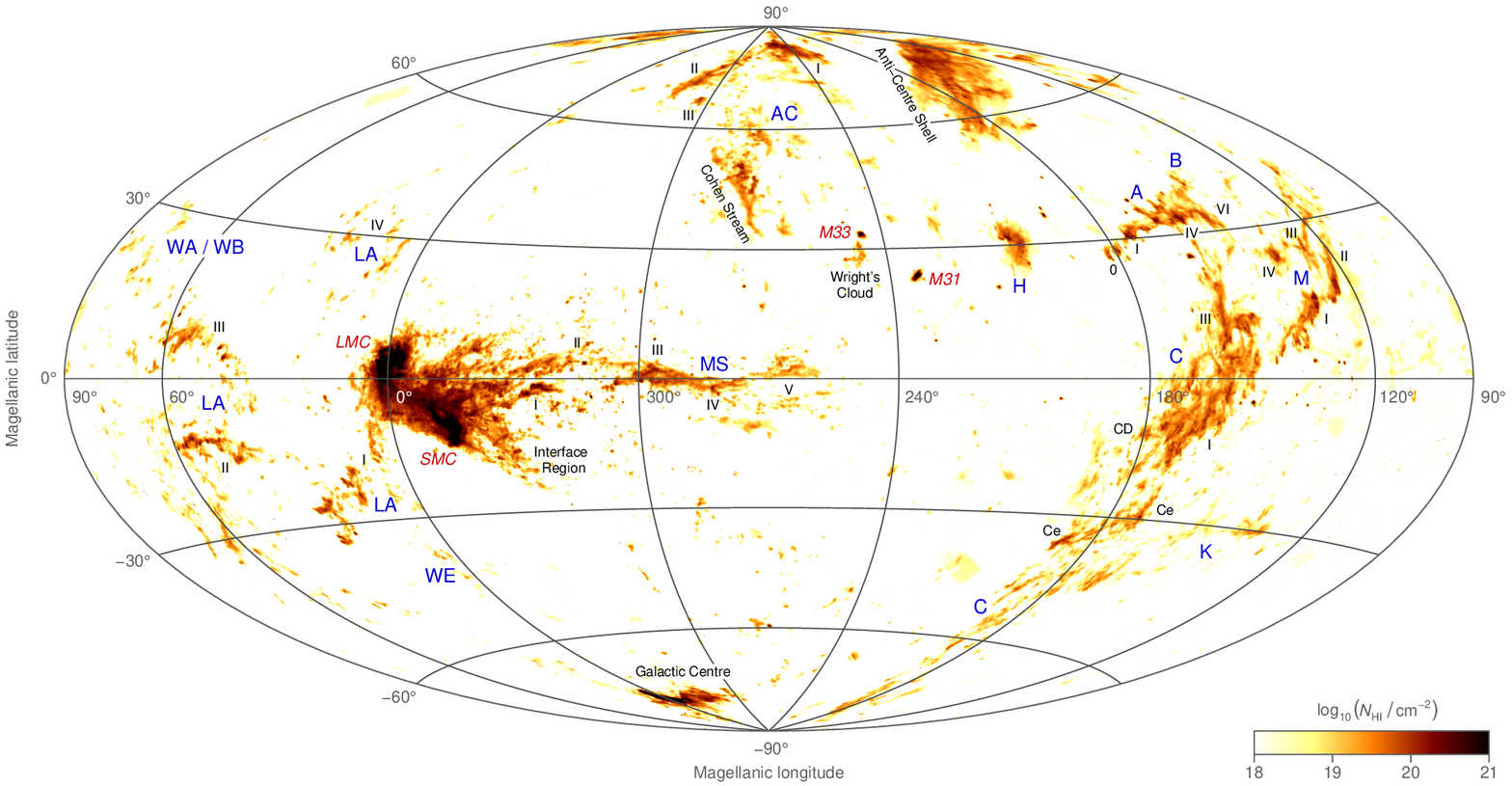}
	\includegraphics[width=\linewidth]{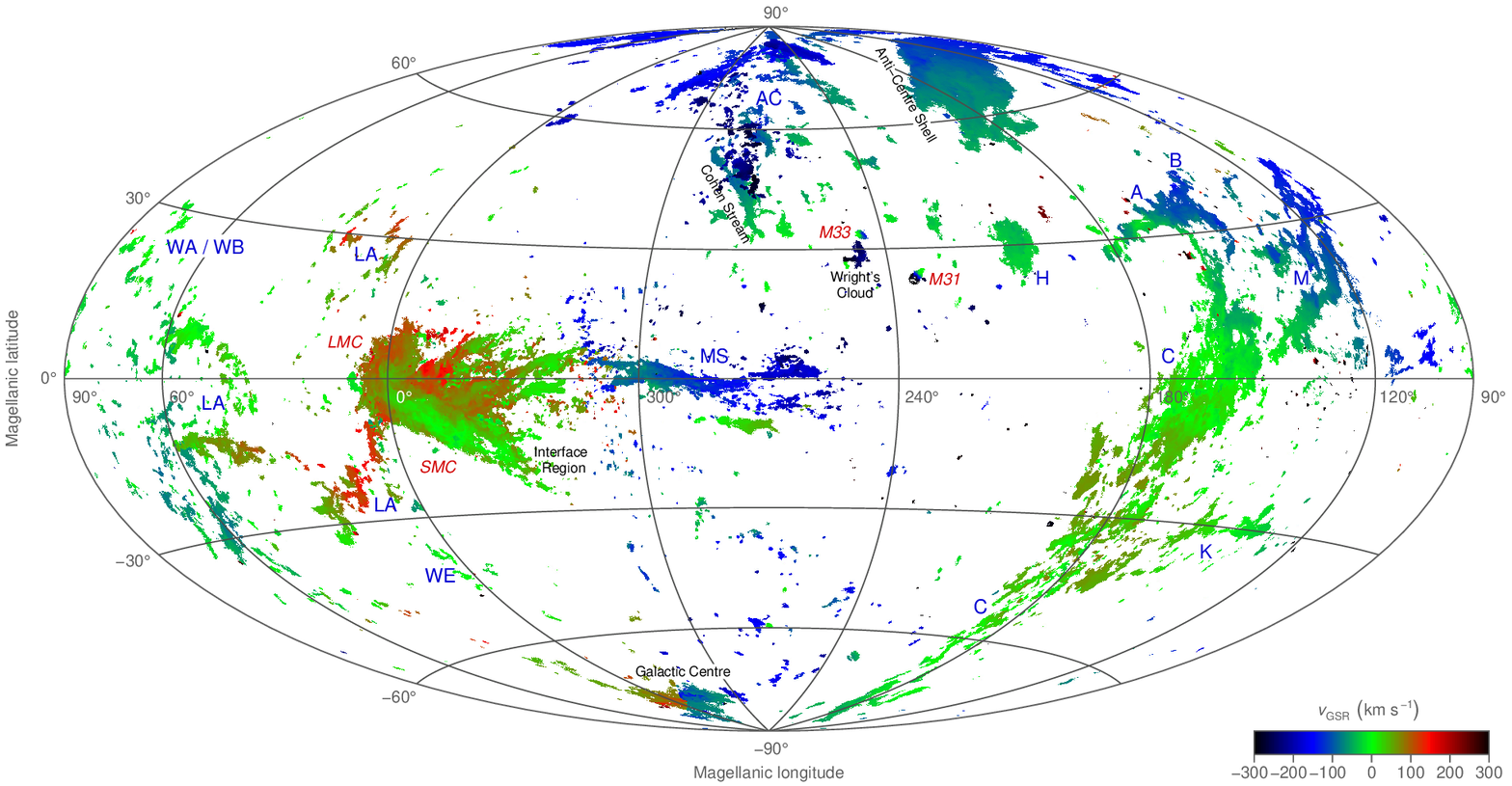}
	\caption{Same as in Fig.~\ref{fig:all-sky-maps}, but using the Magellanic coordinate system as defined by \citet{Nidever2008}. The coordinate system is chosen such that the Magellanic Stream is aligned with the equator to facilitate studies of the Magellanic system. Note that the map is centred on $270^{\circ}$ in Magellanic longitude, as otherwise complex~C would be wrapped around the edge of the map.}
	\label{fig:all-sky-maps-2}
\end{figure*}


\bsp	
\label{lastpage}
\end{document}